\documentclass[useAMS]{mn2e}
\usepackage{graphicx}
\usepackage{amssymb}

\title [Dark Sector through a bounding curvature criterion] {Approaching the Dark Sector through a bounding curvature
  criterion}

\author[X. Hernandez, R. A. Sussman and L. Nasser] {X. Hernandez$^{1}$, R. A. Sussman$^{2}$ and L. Nasser$^{3}$\\ 
$^{1 }$ Instituto de Astronom\'{\i}a, Universidad Nacional Aut\'{o}noma de M\'{e}xico,
Apartado Postal 70--264 C.P. 04510 M\'exico D.F. M\'exico. \\
$^{2}$ Instituto de Ciencias Nucleares, Universidad Nacional Aut\'{o}noma de M\'{e}xico,
Apartado Postal 70--543 C.P. 04510 M\'{e}xico  D.F., M\'{e}xico.\\
$^{3}$ Department of Science and Mathematics, Columbia College, Chicago, 1L 60605, USA.
}

\date{Released 13 September 2018}

\pagerange{\pageref{firstpage}--\pageref{lastpage}} \pubyear{2017}

\begin{document}

\label{firstpage}

\maketitle

\begin{abstract}
  Understanding the observations of dynamical tracers and the trajectories of lensed photons at galactic scales within
  the context of General Relativity (GR), requires the introduction of a hypothetical dark matter dominant component.
  The onset of these gravitational anomalies, where the Schwarzschild solution no longer describes observations, closely
  corresponds to regions where accelerations drop below the characteristic $a_{0}$ acceleration of MOND, which occur at a well
  established mass-dependent radial distance, $R_{c}\propto (GM/a_{0})^{1/2}$. At cosmological scales, inferred dynamics are also
  inconsistent with GR
  and the observed distribution of mass. The current accelerated expansion rate requires the introduction of a hypothetical dark
  energy dominant component. We here show that for a Schwarzschild metric at galactic scales, the scalar curvature, K, multiplied
  by $(r^{4}/M)$ at the critical MOND transition radius, $r=R_{c}$, has an invariant value of $\kappa_{B}=K(r^{4}/M)=28Ga_{0}/c^{4}$.
  Further, assuming this condition holds for $r>R_{c}$, is consistent with the full spacetime which under GR
  corresponds to a dominant isothermal dark matter halo, to within observational precision at galactic level. For a FLRW metric,
  this same constant bounding curvature condition yields for a spatially flat spacetime a cosmic expansion history which agrees with the
  $\Lambda$CDM empirical fit for recent epochs, and which similarly tends asymptotically to a de Sitter solution. Thus, a simple
  covariant purely geometric condition identifies the low acceleration
  regime of observed gravitational anomalies, and can be used to guide the development of {  extended} gravity theories at both
  galactic and cosmological scales.

\end{abstract}


\section{Introduction} \label{intro}

Within GR, the gravitational anomalies detected at galactic scales are viewed as indirect evidence for a
dominant dark matter halo. Such an approach is generally consistent with observations, and leads to the $\Lambda CDM$
paradigm at galactic and cosmological scales. However, over the last years alone, LHC results 
eliminating simple super-symmetric candidates (CMS collaboration 2016), the astrophysical searches for dark matter
annihilation signals being consistent with no dark matter signal (e.g. Fermi-LAT and DES collaborations 2016
searching for dark matter annihilation signals in local dwarf galaxies reporting results consistent with expected
backgrounds), and various recent direct detection experiments returning only ever stricter exclusion limits (e.g.
Yang et al. 2016 yielding no dark matter signal from the PANDAX-II experiment, ruling out previous claims, and
Szydagis et al. 2016 for the LUX and LZ collaborations returning also zero dark matter signal), encourage the
sustained search for {  extended theories of gravity} where the need for dark, undetected, hypothetical components might
be eliminated.

The first such attempt to identify that 'dark matter' gravitational anomalies occur only (and always) at acceleration
scales below a critical value was MOND, Milgrom (1983). Under MOND and MONDian theories, the principal features
are a change in gravity for acceleration scales below $a_{0}=1.2 \times 10^{-10} m/s^{2}$, where centrifugal equilibrium
velocities (and velocity dispersion for pressure supported systems to within a factor of order unity) about a spherically
symmetric mass $M$, become constant at a value of $V_{TF}=(G M a_{0})^{1/4}$.

MONDian dynamics have been reported at scales of wide binary stars (Hernandez et al. 2012, Hernandez et al. 2018), Galactic
globular clusters (e.g. Scarpa et al. 2003, Hernandez et al. 2017), local dwarf spheroidal galaxies
(e.g. Hernandez et al. 2010, L\"{u}ghausen et al. 2014), elliptical galaxies (e.g. Jim\'enez et al. 2013,
Tian \& Ko 2016), spiral galaxies (e.g. McGaugh \& de Blok 1998, Lelli 2016) and the overall velocity dispersion
profiles and radial acceleration relation across over 7 orders of magnitude in mass for galactic systems (e.g.
Dabringhausen et al. 2016, Lelli et al. 2017, Durazo et al. 2018).

In MOND and similar theories (e.g. Bekenstein 2004,  Moffat \& Toth 2008, Zhao \& Famaey 2010, Capozziello \& De Laurentis 2011,
Mendoza et al. 2013) the transition in regimes for gravity is generally introduced by hand, with no clear
fundamental physical principle identified as the source of this regime transition. Further, the identification at low velocities
of a critical transition acceleration as the MOND $a_{0}$ constant seems problematic in terms of constructing  fundamental
underlying covariant theories, where acceleration always appears as frame dependant. At cosmological scales, the numerical
coincidence of $a_{0} \approx c H_{0}$ also remains unexplained, with a relation between $a_{0}$ and the cosmological constant
sometimes being suggested (e.g. Verlinde 2016). We here show that a purely geometric condition, the product of the Kretschmann
curvature scalar and $(r^{4}/M)$ having a fixed minimum value, identifies the transition to the low acceleration MONDian regime.
Moreover, keeping this condition fixed beyond this transition is consistent with the empirically determined metric coefficients
identified with a dominant dark matter halo under GR and Newtonian physics.

Lastly, it is remarkable that the same geometric condition imposed upon a FLRW flat universe yields an asymptotically de
Sitter universe with an expansion rate accurately tracing the $\Lambda$CDM concordance fit for recent epochs. It is therefore
suggested that the geometric condition presented here provides important clues as to the origin of the regime transition for
gravity towards low accelerations, and must be satisfied by any extended theories of gravity attempting to explain observed
gravitational anomalies in the absence of dark {\it ad hoc} entities.

Section (2) shows the invariance of the product of the Kretschmann scalar times $(r^{4}/M)$ when
evaluated at the MOND transition radius, $r=R_{c}$. Extending this $\kappa_{B} = K(r^{4}/M) = const.$ condition to $r>R_{c}$ under a
spherically symmetric and static metric is shown to be consistent with the metric potentials required
by rotation curve and lensing inferences, this time in the absence of any dark matter, in section (3). In
section (4) we explore the cosmological implications of the constant bounding curvature criterion by applying
it to a flat FLRW universe, and obtain an asymptotic de Sitter solution compatible with current
observations. Our conclusions appear in section (5).

\section{Bounding curvature invariance at the MOND transition radius for Schwarzschild spacetime}

As is well known, the gravitational anomalies often attributed to the presence of a dominant dark matter
halo at galactic scales, appear always (and only), in the low acceleration regime, where the acceleration drops
below the critical MOND acceleration constant of $a_{0}=1.2 \times 10^{-10} m/s^{2}$ e.g. Milgrom (1984), Famaey \&
McGaugh (2012). This occurs at a critical mass-dependent radius given by:

\begin{equation}
R_{c}=\alpha R_{M}= \alpha \left( GM/a_{0} \right)^{1/2},
\end{equation}

\noindent where $M$ is the total baryonic mass of a galactic system, $R_{M}=(GM/a_{0})^{1/2}$ is the MOND radius of the system
in question and $\alpha$ is a dimensionless number of order unity. In general, the baryonic mass has essentially converged by
$R_{c}$, with any remaining baryonic matter
distribution beyond being a very minor component usually treated as constituting test particles. 
In the high acceleration regime, where $a>a_{0}$, no gravitational anomalies
appear beyond observational errors and uncertainties in inferred baryonic mass to light ratios. Hence, whenever
$r<R_{c}$ the validity of GR is unchallenged. Thus, we can assume the validity of the Schwarzschild metric:

\begin{eqnarray}
ds^{2}=dr^{2}\left(1-\frac{2 G M}{c^{2} r} \right)^{-1}+r^{2}\left( d\theta^{2}+sin^{2}\theta d\varphi^{2} \right) \nonumber \\
-c^{2}dt^{2}\left(1-\frac{2 G M}{c^{2} r}\right), 
\end{eqnarray}

\noindent for the radial range $r<R_{c}$. Beyond this point, covariant extended theories of gravity attempting to
explain observed dynamics of massive and massless particles in the absence of dark matter, impose a change in regime
to some modified version of GR, e.g. TeVeS of Bekenstein (2004),
f(X) in Mendoza et al. (2013), conformal gravity in Mannheim \& Kazanas (1994). Generally, this change in regime
is introduced by hand, with no fundamental physical explanation for a change in regime for gravity at this mass-dependent
radius of $R_{c}$. Relative exceptions are the recent works of McCulloch (2017) and Verlinde (2016)
where however, it is the effect of a cosmological constant resulting from a quantum vacuum which leads to $R_{c}$.
It is our interest to attempt a geometric explanation for the entirety of the dark sector, so explaining galactic
dark matter as a result of cosmological dark energy appears unsatisfactory. It is interesting that a purely geometric
invariant appears for all masses at $R_{c}$, as shown below.

{  We here attempt to construct a covariant description of the problem grounded on observations of galactic dynamics, rather
than starting from proposing a covariant extension to GR from which testable dynamical consequences in the low velocity regime
then follow, e.g. Capozziello et al. (2007). Thus, we present a complementary approach built on identifying curvature invariants
associated with empirical inferences. The first choice might be the Ricci scalar, which however is of limited use as in the
$R<R_{c}$ GR regime, it is cero for the Schwarzschild metric. We turn to the Kretschmann curvature scalar:}

\begin{equation}
K=R_{\mu \nu \alpha \beta} R^{\mu \nu \alpha \beta},
\end{equation}

\noindent a full contraction of the $R^{\mu \nu \alpha \beta}$ Riemann tensor, which under a Schwarzschild metric has a
value given by:

\begin{equation}
K_{S}=48 \frac{G^{2} M^{2}}{c^{4} r^{6}}.
\end{equation}

Notice that multiplying by $r^{4}/M$, and evaluating the resulting
quantity at the critical radius of $R_{c}$, we obtain a constant value for all masses:

\begin{equation}
K_{S}\left(r^{4}/M \right)|_{r=R_{c}} = 48 \left( \frac{Ga_{0}}{\alpha^{2}c^{4}} \right).
\end{equation}

We see that a quantity which we shall henceforth refer to as bounding curvature, $\kappa_{B}=K(r^{4}/M)$, has a fixed
value for all masses at the transition point beyond which gravitational anomalies appear at galactic systems, the
region where extended theories of gravity introduce modifications to GR. We see from equation (4) that $\kappa_{B}$
decreases radially with $r^{-2}$ within a Schwarzschild spacetime, until reaching always a fixed value at
$r=R_{c}$, for all masses. 

{  The existence of a second gravitational mass dependent scale radius (in addition to the Schwarzschild radius) is expected
in the context of f(R) extended theories of gravity. For example, in Capozziello \& de Laurentis (2011) and
Capozziello et al. (2017) it is shown that requiring the existence of Noether symmetries in a general f(R) theory, implies
the appearance of a second critical radius and an asociated geometric invariant. If one further requires a MONDian weak field
limit, this extra gravitational radius naturally yields $R_{c}$. Our approach is complementary, as we are showing explicitly
what the geometric conserved quantity asociated to $R_{c}$ should be.}

\section{Consistency of constant bounding curvature with inferred spacetime beyond $R_{M}$ at Galactic scales}


We now explore the extension of the $\kappa_{B}=K (r^{4}/M)$ condition to the radial range $r>R_{c}$, where
isothermal dark matter halos are invoked to explain observed dynamics and lensing observations under GR.
Working under a general spherically symmetric and static metric,

\begin{equation}
ds^{2}=B(r) dr^{2} +r^{2}\left(d\theta^{2}+sin^{2}\theta d\varphi^{2} \right) -A(r)c^{2}dt^{2}.
\end{equation}

The Kretschmann scalar curvature is now given by the following expression:

\begin{eqnarray}
K_{SSS}=\frac{A_{r,r}}{(AB)^{3}}\left[ A_{r,r}AB-A_{r}AB_{r} - A_{r}^{2}B\right]  \nonumber \\
+\frac{A_{r}^{2}}{4(AB)^{4}} \left[B_{r}^{2} A^{2} + 2A_{r}B_{r}AB + (A_{r}B)^{2} + 8 \left( \frac{AB}{r}\right)^{2} \right] \nonumber\\
+ \frac{2}{(Br)^{4}} \left[(B_{r}r)^{2}+2B^{4}-4B^{3} +2B^{2}\right].
\end{eqnarray}

The consistency of the constant bounding curvature condition in the 'dark halo' region can be tested by introducing
the metric potentials of the isothermal dark matter halo into the above expression. These are:

\begin{equation}
A=1+2\beta^{2}ln(r/R_{M})\simeq \left(r/R_{M} \right)^{2\beta^{2}}, \hfill B=1+2\beta^{2},
\end{equation}

\noindent where $\beta=V_{TF}/c$, $V_{TF}=(G M a_{0})^{1/4}$ the Tully-Fisher flat asymptotic galactic
rotation curve of a system of baryonic mass $M$ e.g. Misner, Thorne \& Wheeler (1973) section (25.7), Campigotto et al. (2016).

The above empirical metric potentials reduce equation(7) above to:

\begin{equation}
K_{SSS}=28 \left( \frac{\beta}{r} \right)^{4}  +\mathcal{O}\left(\beta^{6}/r^{4}\right).
\end{equation}

\noindent

{  As in the galactic case $V_{TF} \lesssim 300 km/s$, $\beta $ will always be $\lesssim 10^{-3}$. Therefore, the corrections to the
leading term in equation (9) will be at most $10^{-6}$ times smaller than the leading one.}
Hence, droping terms of order $\beta^{6}$ and higher, multiplying the above by $r^{4}/M$, the constant bounding curvature
condition $\kappa_{B}=K (r^{4}/M)$ for the spacetime generally identified with hypothetical isothermal dark matter halos reads:

\begin{equation}
\kappa_{B}=28 \frac {(Ga_{0})}{c^{4}},
\end{equation}

\noindent which will be identical to the value of this parameter at the critical radius for transition towards
the modified regime of eq.(5) provided:

\begin{equation}
\alpha=\left( \frac{48}{28}  \right)^{1/2} = 2\left( \frac{3}{7} \right)^{1/2} =1.309.
\end{equation}

\begin{figure}
\includegraphics[width=9.0cm,height=7.0cm]{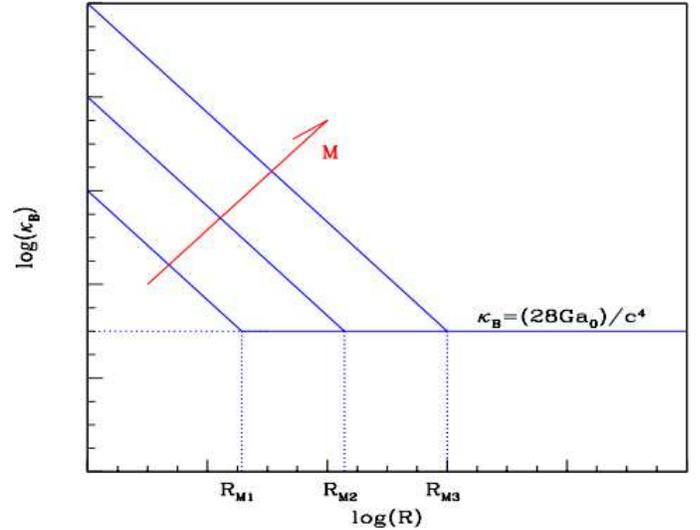}
\caption{The figure shows schematically the $\kappa \propto M/r^{2}$ scaling over the approximately Schwarzschild region
of galaxies, such that on reaching $r=R_{c}=2(3GM/7a_{0})^{1/2}$ for different galaxies with masses ($M_{1}<M_{2}<M_{3}$),
a fixed value of $\kappa_{B}$ results. Demanding this fixed value remains constant for $r>R_{c}$ yields a spacetime
consistent with the isothermal dark matter halos inferred under GR to extend around observed galaxies.}
\end{figure}

Figure (1) gives an schematic representation of the results discussed so far; within the approximately Schwarzschild region
$\kappa_{B}$ scales as $\propto (M/r^{2})$, such that on reaching the critical transition radius of $R_{c}=2(3/7)^{1/2}R_{M}=
2(3/7)^{1/2}(GM/a_{0})^{1/2}$ relevant for
each galaxy, $\kappa_{B}$ has always the same constant value. We see that the spherically symmetric and static spacetime
of the isothermal dark matter halos inferred under GR from observed kinematics and lensing observations of galaxies,
{  is consistent with} a fixed value of $\kappa_{B}$ beyond $R_{c}$.

Thus, a purely geometric quantity with units of $M^{-1}$ appears,
which falls with radius until reaching a minimum value at $R_{c}$, at which it remains fixed, at least for the region currently
sampled by kinematic and lensing galactic observations. {  The scales over which this ideas apply can be extended to galaxy cluster
scales, through comparing with the results of Capozziello et al. (2009) and Capozziello et al. (2018), where the existence of a
critical radius of order $R_{c}$ leading to a full description of dynamics in galaxy clusters within a f(R) framework, is presented.}

We see that through the geometric condition $\kappa_{B}=28Ga_{0}/c^{4}$ the acceleration constant of MOND can be defined in terms
of a purely geometric condition for a critical value of $\kappa_{B}$ having units of $kg^{-1}$. Also, it is easy to show through
dimensional analysis that any other product of powers of $K$, $r$ and $M$ which satisfies having both a constant value at $r=R_{c}$
and that same fixed value consistent with equation (8) for $r>R_{c}$, will necessarily be a power of the condition introduced,
for which we keep $\kappa_{B}=K(r^{4}/M)$.

\section{Consistency of constant bounding curvature with inferred spacetime at cosmological level}

We now test the bounding curvature criterion presented above in a cosmological context, where matter-energy sources at
sufficiently large scales are described as perturbations on a homogeneous and isotropic FLRW background whose metric is given by:

\begin{equation}
ds^{2}=dr^{2} \left( \frac{a^{2}(t)}{1-kr^{2}}  \right) +a^{2}(t)r^{2}(d\theta^{2}+sin^{2}\theta d\varphi^{2})-c^{2}dt^{2}.
\end{equation}

The Kretschmann invariant now becomes:

\begin{equation}
K_{F}=\frac{12}{(ca)^{4}} \left[ (a \ddot{a})^{2}+(k-\dot{a}^{2})^{2}   \right],
\end{equation}

\noindent where dot denotes temporal derivatives. Taking a spatially flat universe $k=0$ yields,

\begin{equation}
K_{F}=\frac{12}{c^{4}} \left[ (\dot{H}+H^{2})^{2} +H^{4}  \right]  
\end{equation}

\noindent where we have introduced the Hubble scalar, $H=\dot{a}/a$. We now multiply the expression for $K_{F}$ above
by $(R_{F}^{4}/M)$, where we take $M$, assumed as a constant, as the mass within a sphere of radius $(c/H_{0})$ and
density $0.04\rho_{c}$, with
$\rho_{c}=3 H_{0}^{2}/(8\pi G)$, i.e., the locally inferred total baryonic matter density of the universe. $R_{F}$ now
becomes the natural characteristic length-scale furnished by the Hubble radius $\gamma (c/H)$, with $\gamma$ a dimensionless
constant expected to be of order unity, and which includes all our uncertainty in going from the spherically symmetric and
static spacetime of galactic dynamics, to the homogeneous, isotropic and dynamic case of a FLRW universe. {  This yields:}

\begin{equation}
\kappa_{B}= \frac{24 \gamma^{4} G H_{0}} {0.04 c^{3}} \left[ \left( \frac{\dot{H}}{H^{2}} +1 \right)^{2} +1   \right].
\end{equation}

Notice that in using only the Hubble scalar, $c/H$ and $c/H_{0}$ in the estimates of characteristic radius and mass of
the system, the geometric condition of equation (10) has become a fully covariant expression in equation (15).
The constant bounding curvature condition $\kappa_{B}=28 Ga_{0}/c^{4}$ in the cosmological context
becomes:

\begin{equation}
\left[ \left( \frac{\dot{H}}{H^{2}} +1\right)^{2} +1 \right]= \frac{8\times 10^{-3}}{\gamma^{4}},
\end{equation}

\noindent where we have used the well known empirical numerical relation $c H_{70}=5.83 a_{0}$ (e.g. Milgrom 2002).
The above equation can be readily integrated to yield:

\begin{equation}
\frac{H}{H_{0}}=\left[1+X^{\pm}(T-T_{0})   \right]^{-1},
\end{equation}

\begin{equation}
a(T)=\left[X^{\pm}(T-T_{0}+1)   \right]^{1/X^{\pm}}.
\end{equation}

In the above $T=tH_{o}$, $T_{0}=t_{0}H_{0}$ and $X^{\pm}=1-\pm(8\times10^{-3}/\gamma^{4}-1)^{1/2}$, where the present time is
denoted by $t_{0}$. From eq.(17) we see that imposing the condition $\kappa_{B}=28Ga_{0}/c^{4}$ on a flat FLRW universe
results in an asymptotically de Sitter universe with one free parameter, $\gamma$, the scaling between $c/H$ and the relevant
scale radius for the system.

\begin{figure}
\includegraphics[width=9.0cm,height=7.0cm]{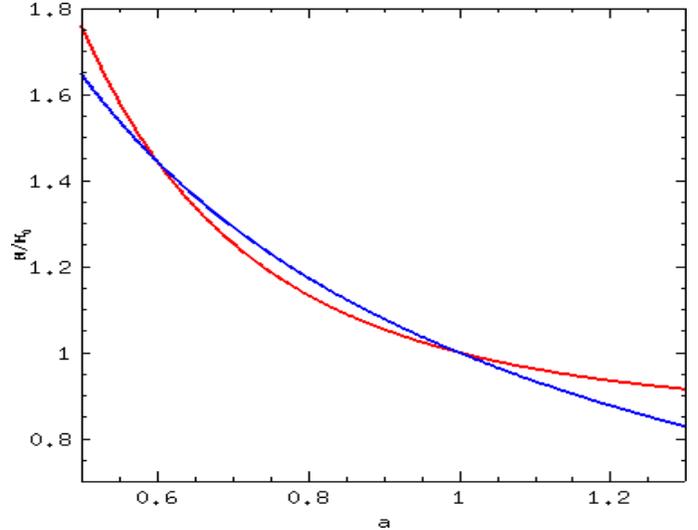}
\caption{The figure shows the recent cosmological expansion history through the Hubble constant in units
of its present value, as a function of the scale factor.
The upper curve to the right of the plot gives the $\Lambda$CDM cosmology, while the lower one to the right of the plot
shows the exact analytical results for equation (17), the constant bounding curvature condition imposed upon a spatially flat FLRW
universe.}
\end{figure}

We can now try to find an optimal value for $\gamma$ by comparing equations (17) and (18) to the empirically inferred
expansion history of the concordance $\Lambda$CDM universe, this last given by:

\begin{equation}
a(t)=\left(\frac{\Omega_{M}}{\Omega_{\Lambda}}\right)^{1/3}\left[ sinh\left(\frac{3}{2}\sqrt{\Omega_{\Lambda}}H_{0}t\right)\right]^{2/3},
\end{equation}

\noindent for parameters $\Omega_{M}=0.3$ and $\Omega_{\Lambda}=0.7$. The corresponding expansion history is plotted in
figure (2) as a function of time in units of $T=H_{0}t$,  upper curve to the right of the plot. The other
curve in the figure is a solution to equations (17) and (18) for $\gamma=0.2935$ and taking the $+$ sign, i.e., a very reasonable
choice of characteristic radius of $R_{F}=0.2935(c/H)$. We see that plotting both in terms of their respective $H/H_{0}$ values, by
construction, both curves meet at $a=1$. However, it is extremely encouraging of the ideas presented, that over the whole $1/2<a<1$
range plotted, both curves remain within $5\%$ of each other. As present uncertainties in the determination of $H_{0}$ are of this order,
and since the value of $H(a)$ is necessarily less well constrained at higher redshifts, it is clear that the model presented reproduces
the observed recent accelerated expansion history of the universe, in this case, as a result of a fully covariant geometric condition, and
without invoking dark energy. Thus, an alternative theory of gravity that incorporates the geometric condition (17) would imply a
conceptually different approach to cosmic dynamics not requiring any hypothetical dark energy components to yield a recent accelerated
expansion phase.


It might appear counter intuitive that the same condition which allows to explain observed rotation curves in the
absence of dark matter should also yield cosmic accelerated expansion in the absence of dark energy, since at a naive
zero order level, galactic dark matter provides ``extra attractive'' gravity, while cosmological dark energy leads
to ``repulsive'' gravity. The answer lies in considering the very different boundary and symmetry conditions of both limits;
galactic rotation curves arise in bound systems that are approximately isolated and well described by spherically symmetric
and static metrics that are approximately asymptotically flat, and therefore where one can identify  
a well defined centre towards which gravity pulls, and thus, extra gravity leads to more pull.
At cosmological scales matter--energy can no longer be described as stationary near asymptotically flat localised sources,
but rather as close to homogeneous patches that expand isotropically and are all statistically equivalent, thus together
approximating an FLRW model. All observers see others receding, being pulled away faster in all directions by the same
geometric modification to gravity, yielding a different effect for the case of a very distinct assumed symmetry.

This is in fact what has been shown to occur
with some $f(\chi)$ {  extended} gravity options tailored to reproduce observed galactic dynamics in the absence of
dark matter (Mendoza et al. 2013), which also yield agreement with the observed SN Hubble plot (Carranza et al. 2013).
{  Both metrics treated here} are idealisations, we have shown that the proposed geometric condition is consistent with empirical
inferences for spacetimes at both limits. Indeed, it is tempting to speculate that the covariant geometric $\kappa_{B}=const.$
condition might arise from some more fundamental holographic or entropic criterion, and perhaps form the basis of
a more complete extended gravity.



\section{Concluding remarks}

We have shown that for a Schwarzschild metric, the mass dependant MOND transition radius corresponds to an
invariant purely geometric condition given by the bounding curvature $\kappa_{B}=K(r^{4}/M)=28Ga_{0}/c^{4}$,
with $\kappa_{B}$ having units of $kg^{-1}$. {  This hence identifies the specific geometric invariant associated
to $R_{c}$, once this length scale becomes the second mass dependent gravitational radius which arises in f(R)
theories having imposed a Noether symmetry, e.g. Capozziello et al. (2017).} Thus, the often mentioned incompatibility
of the MOND 'acceleration scale' with covariant frameworks is {  in fact not present}.

Assuming a spherically symmetric and static metric for $r>R_{c}$ and imposing $\kappa_{B}=const.$, is compatible with the
empirically determined metric potentials usually interpreted as evidence for a dominant dark matter halo, without the
need of any dark matter.

Imposing the $\kappa_{B}=28Ga_{0}/c^{4}$ condition on a flat FLRW metric results in an asymptotic de Sitter
solution which for recent epochs ($z<1$) accurately traces the accelerated concordance $\Lambda$CDM empirical fit to better than
$5 \%$ without the need of invoking any dark energy.

The condition found can be used as a necessary restriction on {extended} theories of gravity seeking to explain
the observed gravitational anomalies in absence of hypothetical {and undetected} dark matter and dark energy components.
Indeed one can speculate that a fixed minimum $\kappa_{B}$ principle could constitute the basis of a covariant modified
theory of gravity.

\section*{acknowledgements}

XH acknowledges support from DGAPA-UNAM PAPIIT IN-104517 and CONACyT, and RAS acknowledges
support from CONACYT 239639 and PAPIIT-DGAPA RR107015.

\end{document}